\documentclass[11pt,pra,nofootinbib,superscriptaddress,floatfix]{revtex4}

\usepackage{graphics}           
\usepackage{bm}                 
\usepackage{amsmath}            
\usepackage{amsfonts}
\usepackage{amssymb}
\usepackage{latexsym}           

\small\normalsize

\newcommand{\ket}[1]{\mbox{$|#1\rangle$}}

\newcommand{\identity}{\leavevmode\hbox{\small1\kern-3.2pt\normalsize1}}

\begin{document}

\title{Decoherence vs entanglement in coined quantum walks}

\author{Olivier Maloyer}
\affiliation{Magist{\`e}re de Physique Fondamentale d' Orsay, Universit\'{e} Paris-Sud, Orsay,   France}
\affiliation{School of Physics and Astronomy, University of 
Leeds, LS2 9JT, United Kingdom.}

\author{Viv Kendon}
\email{V.Kendon@leeds.ac.uk}
\affiliation{School of Physics and Astronomy, University of 
Leeds, LS2 9JT, United Kingdom.}

\date{December 29th, 2006, revised March 11th, 2007}

\begin{abstract}
Quantum versions of random walks on the line and cycle show a quadratic
improvement in their spreading rate and mixing times respectively.
The addition of decoherence to the quantum walk produces a more uniform 
distribution on the line, and even faster mixing on the cycle by
removing the need for time-averaging to obtain a uniform distribution.
We calculate numerically the entanglement between the coin and the
position of the quantum walker and show that the optimal decoherence
rates are such that all the entanglement is just removed by the time
the final measurement is made.
\end{abstract}


\maketitle

\tableofcontents


\section{Introduction and background}
\label{sec:intro}

Simple quantum generalisations of classical random walks 
spread quadratically faster on the line \cite{ambainis01a,nayak00a},
and mix quadratically faster on the cycle \cite{aharonov00a}.
These promising examples of a quantum speed up were soon followed by 
several algorithms based on quantum walks. 
Shenvi et al.~\cite{shenvi02a} proved a quantum walk
can solve the unsorted database search problem quadratically faster,
and Childs et al.~\cite{childs02a} proved an exponential 
speed up for crossing a particular type of graph.  Several more
algorithms have followed these, Ambainis
\cite{ambainis03a} gives an overview of quantum walk algorithms, and
Kempe \cite{kempe03a} provides an introductory review
of quantum walks and their properties.
Physical implementations of quantum walks have also been proposed
\cite{travaglione01a,sanders02a,dur02a}, as sensitive tests of
coherent control over quantum particles.
Optical implementation dates back to Bouwmeester et al.~\cite{bouwmeester99a},
which can be viewed as a quantum walk on the
line with many photons \cite{knight03a,kendon04a}.
A quantum walk on a cycle with four nodes has been implemented on a
NMR (nuclear magnetic resonance) quantum computer using three qubits,
two for the position label (in binary) and one for the coin \cite{ryan05a}.

Quantum walks on simple one-dimensional structures remain a fertile
testing ground for further research.
It was shown numerically by Kendon and Tregenna \cite{kendon02c}
that the addition of decoherence or measurements to the quantum walk
dynamics can optimise the spreading and mixing properties for quantum walks on
both the line and the cycle
(recently proved by Richter \cite{richter06a,richter06b}).
A detailed survey of the effects of 
decoherence in quantum walks can be found in \cite{kendon06b}.
In this work we look closely at how the interplay between quantum evolution
and decoherence or measurements produces optimal computational properties.
In particular, we calculate the entanglement between the coin and
the position of the quantum walker and observe how it varies as the
quantum walk evolution unfolds.
This study is restricted to quantum walks taking place in discrete
time and space, using a quantum coin to control the choice of direction.
Thus we can use the entanglement between the coin and position
as an indication of how ``quantum'' the walk is as it progresses.
A different method for assessing ``quantumness'' would be needed for
continuous-time quantum walks, as introduced for quantum algorithms by
Farhi and Gutmann \cite{farhi98a}.

In the following subsections we describe the background material:
simple quantum walks, the decoherence model we use, and the
entanglement measure (negativity) we employed.  The results of
our study are then described in \S\ref{sec:results}, first for 
the line, then for cycles, and finally a summary of the common
elements of our findings.


\subsection{Simple, coined, quantum walks}
\label{ssec:qwalk}

A discrete-time (coined) quantum walk dynamics consists of a quantum
``coin toss'' operation $\mathbf{C}$, followed by a shift operation $\mathbf{S}$
 to move the 
quantum walker to a new position.  These are repeated alternately
for $T$ steps of the quantum walk, and the final position of the
quantum walker is measured.
For quantum walks on the line and the cycle, we have just two
choices of which way to step, so the quantum coin is a two state system.
We write $|x,c\rangle$ for a quantum walker at position $x$ with a coin
in state $c \in \{+1,-1\}$.  For a walk on the line, $x \in \mathbb{Z}$
and for a walk on a cycle of size $N$, we have $x\in\mathbb{Z}_N$.
The full state of the quantum walk $|\Psi(t)\rangle$ at time $t$
can be written as a superposition of terms in each
basis state $|x,c\rangle$,
\begin{equation}
|\Psi(t)\rangle = \sum_{x,c}\psi_{x,c}(t)|x,c\rangle,
\end{equation}
where $\psi_{x,c}(t) \in \mathbb{C}$ and the normalisation 
is $\sum_{x,c}|\psi_{x,c}(t)|^2 = 1$.
When the quantum walk is measured (in the basis just defined),
the probability of finding the quantum walker at
position $y$ with the coin in state $b$ is given by
\begin{equation}
P(y,b,t) = |\langle y,b|\Psi(t)\rangle|^2 = |\psi_{y,b}(t)|^2.
\end{equation}

The coin toss $\mathbf{C}$ and shift $\mathbf{S}$ are defined in terms
of their action on the basis states $|x,c\rangle$,
\begin{equation}
\mathbf{S}|x,c\rangle = |x+c,c\rangle
\end{equation}
\begin{equation}
\mathbf{C}|x,c\rangle = (|x,-c\rangle + c|x,c\rangle)/\sqrt{2}
\end{equation}
One can add more general bias or phase into the coin toss operation, see,
for example, \cite{bach02a}, but this does not greatly change the basic
properties of the quantum walk on a line or cycle, so we will consider
only the unbiased case in this paper.
For a quantum walk starting at position $x=0$ with the coin in a
superposition state $(|{-1}\rangle + i|{+1}\rangle)/\sqrt{2}$,
(where $i=\sqrt{-1}$)
we can write a quantum walk of $T$ steps as
\begin{equation}
|\Psi(T)\rangle = (\mathbf{SC})^T\left\{|0,-1\rangle + i|0,+1\rangle\right\}/\sqrt{2}
\end{equation}
The solution for $|\Psi(T)\rangle$ may be obtained by various methods
such as Fourier analysis \cite{nayak00a} and path counting \cite{ambainis01a},
and has been studied extensively.  The key result 
is that spreading on the line proceeds linearly with the
number of time steps.
We can use the standard deviation $\sigma_Q(T)$ of the probability distribution
to quantify the spreading rate.  For a quantum walk on the line,
\begin{equation}
\sigma_Q(T) = \sum_{x,c}x^2P(x,c,T)\simeq\left(1-\frac{1}{\sqrt{2}}\right)^{1/2}T,
\end{equation}
asymptotically in the limit of large $T$ \cite{nayak00a}.  In contrast, for the
classical random walk, the standard deviation is $\sigma_C(T) = \sqrt{T}$.

On the cycle, we are interested in mixing times rather than
spreading.  Mixing times can be defined in a number of different
ways, we choose the main definition given in \cite{aharonov00a},
\begin{equation}
M(\epsilon)=\min\left\{T\enspace|\enspace\forall\enspace t>T: ||P(x,t)-P_u||_{\text{tv}}<\epsilon\right\}
\label{eq:mixdef}
\end{equation}
where $P_u$ is the limiting  distribution over the cycle, and
the total variational distance (TVD) is defined as
\begin{equation}
||P(x,T) - P_u||_{\text{tv}} \equiv \sum_x|P(x,T) - P_u|.
\label{eq:tvd}
\end{equation}
A classical random walk on the cycle mixes to within
$\epsilon$ of the uniform distribution in time
proportional to $N^2\log(1/\epsilon)$, 
where $\epsilon$ can be chosen arbitrarily small.

Pure quantum walks, on the other hand, do not mix to a stationary distribution.
Their deterministic dynamics ensures they continue to oscillate indefinitely.
There are several ways to obtain mixing behaviour,
first explored by Aharonov et al.~\cite{aharonov00a}.
By defining a time-averaged probability distribution for the quantum walk,
\begin{equation}
\overline{P(x,c,T)} = \sum_{t=0}^{T-1}P(x,c,t)
\label{eq:meanP}
\end{equation}
they proved that $\overline{P(x,c,T)}$ does converge to a stationary
distribution on a cycle, and that on odd-sized cycles the stationary
distribution is uniform.
A mixing time can be defined for $\overline{P(x,c,T)}$,
\begin{equation}
\overline{M}(\epsilon)=\min\left\{T\enspace|\enspace\forall\enspace t>T: ||\overline{P(x,t)}-P_u||_{\text{tv}}<\epsilon\right\}.
\label{eq:meanmixdef}
\end{equation}
and Aharonov et al.~\cite{aharonov00a} proved that, for odd-sized cycles,
$\overline{M}(\epsilon)$ is bounded above by $O(\epsilon^{-3}N\log N)$,
almost quadratically faster (in $N$) than a classical random walk.
Kendon and Tregenna \cite{kendon02c},
observed numerically that $\overline{M}(\epsilon)\sim O(N/\epsilon)$, 
this has been recently confirmed analytically by Richter \cite{richter06b}.

Notice that we pay a price for the time-averaging: the scaling with
the precision $\epsilon$ is now linear instead of logarithmic.
Aharonov et al.~\cite{aharonov00a} provide a fix for this in the form of a
``warm start''.  The quantum walk is run several times, each repetition
starting from the final state of the previous run.  A small number of such
repetitions is sufficient to reduce the scaling of the mixing time
$\overline{M}(\epsilon)$ to logarithmic in $\epsilon$.

Both the quantum walk on the line and the cycle thus provide
a quadratic speed up over classical random walks.
This quadratic speed up does not carry over to all quantum walks on
higher dimensional structures, see, for example,
\cite{mackay01a,tregenna03a,krovi06a,krovi07a}.
It remains an open question how ubiquitous this behaviour really is.

\subsection{Decoherence in quantum walks}
\label{ssec:decmodel}

We will consider decoherence in the form of randomly-occurring uncorrelated
non-unitary events added to the quantum walk dynamics already described.
The evolution of the quantum walk
must now be described using a density operator $\bm\rho(t)$ given by
\begin{equation}
\bm\rho(t+1) = (1-p)\mathbf{SC}\bm\rho(t)\mathbf{C}^{\dag}\mathbf{S}^{\dag}
          + p\sum_j\mathbb{P}_j\mathbf{SC}\bm\rho(t)
                \mathbf{C}^{\dag}\mathbf{S}^{\dag}\mathbb{P}_j^{\dag}.
\label{eq:decdyn}
\end{equation}
Here $\mathbb{P}_j$ is a projection that represents the action of the
non-unitary decoherence events and $p$ is the probability of applying
the decoherence per time step, or, completely equivalent mathematically,
to a weak coupling between the quantum walk system and a
Markovian environment with coupling strength $p$.
For a pure state, the density operator $\bm\rho(t)\equiv |\Psi(t)\rangle
\langle\Psi(t)|$, it thus has the normalisation $\rm{Tr}[\bm\rho(t)]=1$.
For $p=0$, equation (\ref{eq:decdyn}) reduces to the pure quantum walk
described in \S\ref{ssec:qwalk}, and for $p=1$, to a classical random walk.

\begin{figure}
    \begin{center}
	\resizebox{0.75\columnwidth}{!}{\includegraphics{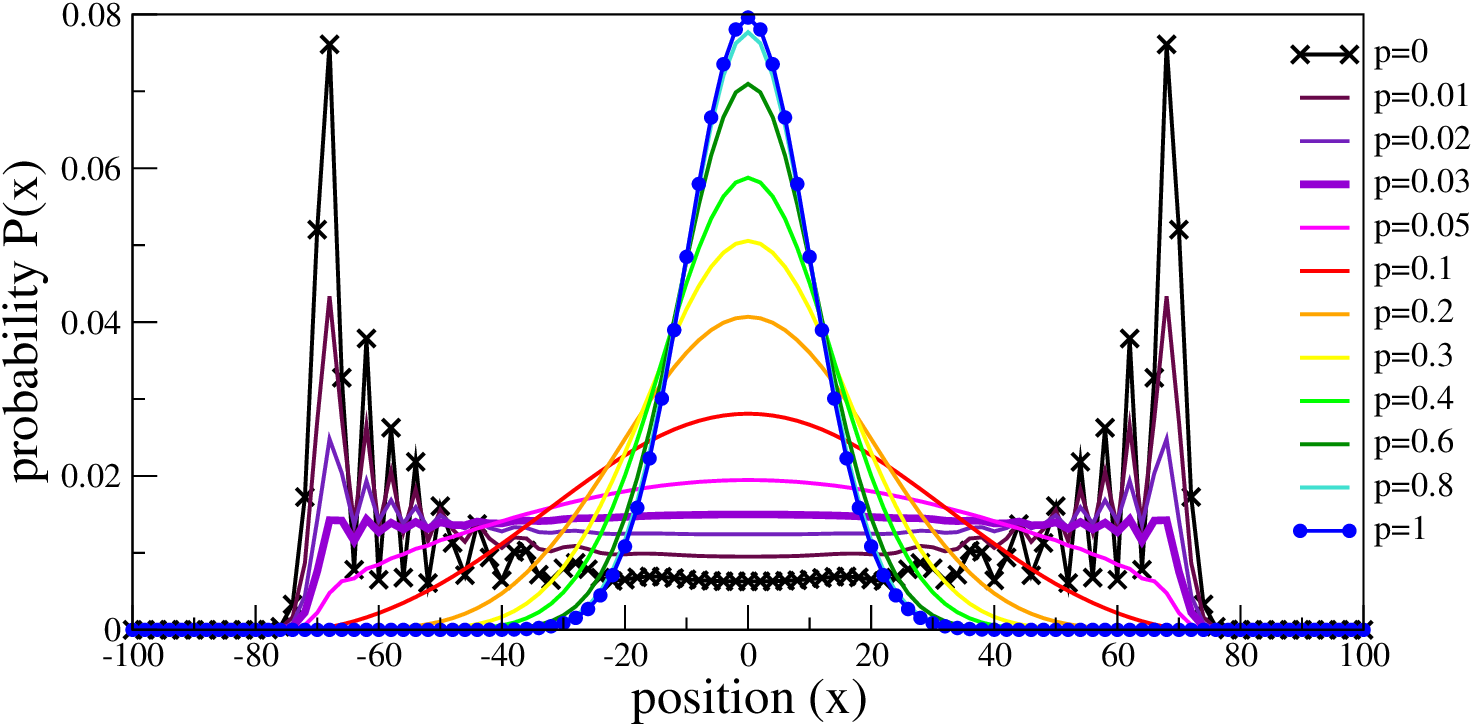}}
    \end{center}
    \caption{Numerical data for a quantum walk on the line of 100 steps,
	with decoherence applied to both the coin and position,
	for various decoherence rates between zero (quantum) and
	one (classical) as shown in the key.}
    \label{fig:pdf100qc}
\end{figure}


Full analytical solution of a non-unitary quantum walk has been
done only for a few special cases.  For quantum walks on the line, Brun et
al.~\cite{brun02c} analysed the case of random measurements on the coin only.
While analytical solution is challenging, equation~(\ref{eq:decdyn})
lends itself readily to numerical simulation since
$\bm\rho$, $\mathbf{S}$ and $\mathbf{C}$ can be manipulated as complex
matrices, and the $\mathbb{P}_j$ generally remove some or all
of the off-diagonal entries in $\bm\rho$.
Kendon and Tregenna \cite{kendon02c} evolved equation~(\ref{eq:decdyn})
numerically for various choices of $\mathbb{P}_j$:
projection onto the position space, projection onto the coin space,
and projection of both coin and position.  In all cases, the spreading
rate is reduced, in the long time limit \cite{brun02c},
it becomes proportional to $\sqrt{T}$ instead of proportional to $T$.
More interesting behaviour is seen for intermediate times
and decoherence rates $p$ with decoherence applied to the position, or 
to both the position and coin.
Kendon and Tregenna observed that, for $2\lesssim pT \lesssim 5$,
the distribution becomes very close to uniform while retaining the
full quantum linear spreading rate, see figure~\ref{fig:pdf100qc}.
With decoherence applied to the coin only, the distribution retains a
cusp shape \cite{kendon02c}.
To quantify the distance from a uniform distribution,
the TVD given by equation~(\ref{eq:tvd}) can be used,
this time with $P_u$ defined to be a top-hat of appropriate width
$x\in\{\pm T/\sqrt{2}\}$, see \cite{nayak00a,ambainis01a}.
Since the quantum walk (in common with the classical random walk on which
it is based) has the property that at odd(even) time steps the walker will
only be found at odd(even)-numbered locations, the top-hat definition of
$P_u$ also incorporates this property.


\begin{figure}
            \begin{center}
            \resizebox{0.6\columnwidth}{!}{\includegraphics{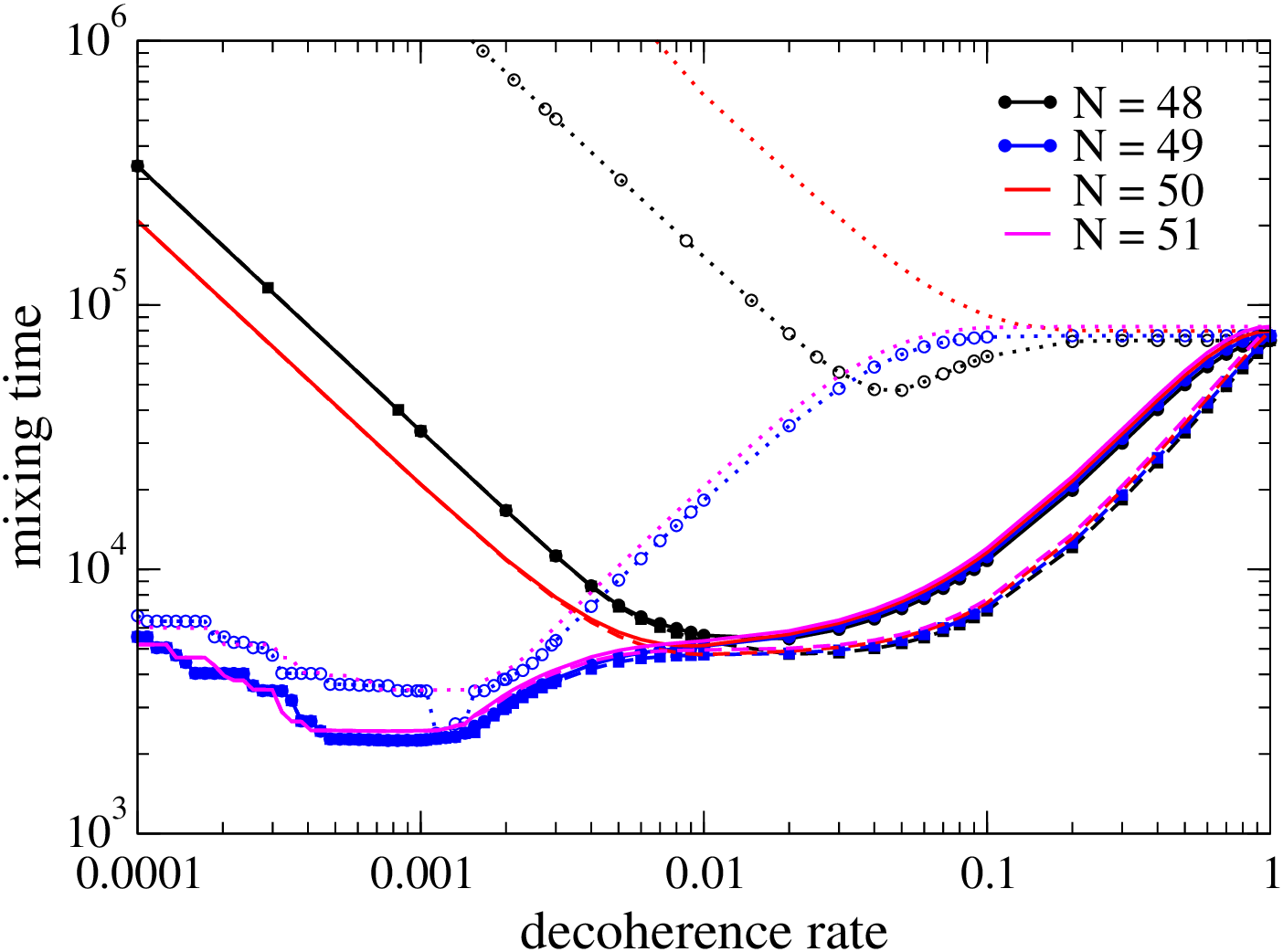}}
            \end{center}
            \caption{Numerical data for the time-averaged mixing time
		$\overline{M}_{\epsilon}(p)$
                on cycles of size $N=48$, $N=49$, $N=50$ and $N=51$,
                for coin (dotted), position
                (dashed) and both (solid) subject to decoherence,
                using $\epsilon=0.002$.}
            \label{fig:c48c49c50c51mixav}
\end{figure}
For a quantum walk on the cycle subjected to decoherence,
the mixing behaviour is dramatically improved provided the decoherence
is applied to the position.  Decoherence guarantees mixing to the
uniform distribution, and a similar judicious choice of
$2\lesssim pN \lesssim 5$ produces a minimum mixing time \cite{kendon02c},
well below the classical value.  Decoherence applied only to the coin
does cause the quantum walk on a cycle to mix,
but not significantly faster than a classical random walk.
Furthermore, time-averaging is no longer necessary, mixing
occurs in time $O(N\log(1/\epsilon)$.  This has recently been proved
by Richter \cite{richter06b}.
\begin{figure}
            \begin{center}
            \resizebox{0.6\columnwidth}{!}{\includegraphics{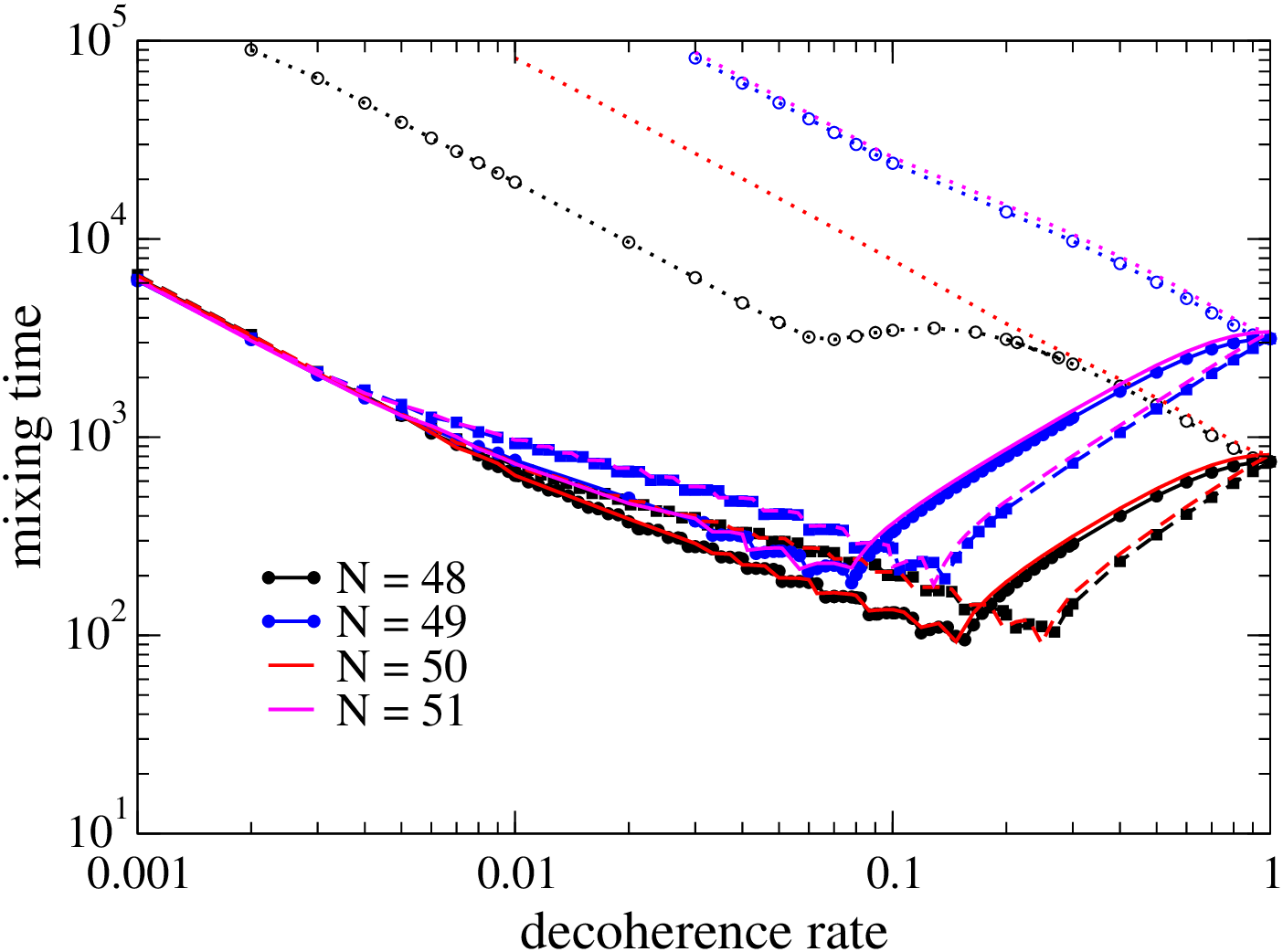}}
            \end{center}
            \caption{Numerical data for the mixing time, without
		time-averaging, $M_{\epsilon}(p)$
		on cycles of size $N=48$ to $N=51$,
                for decoherence applied to coin only (dotted),
                position only (dashed) and both (solid),
                using $\epsilon=0.002$.}
            \label{fig:c48c49c50c51minst}
\end{figure}
This mixing behaviour is illustrated in figures \ref{fig:c48c49c50c51mixav}
and \ref{fig:c48c49c50c51minst} for cycles of size $48$, $49$, $50$ and $51$.  
Figure \ref{fig:c48c49c50c51mixav} shows time-averaged mixing times while
figure \ref{fig:c48c49c50c51minst} shows mixing times without any time
averaging.  Note the scales are different between the two figures, the
mixing times are longer when time-averaging is applied, reflecting the
change from logarithmic to linear dependence on $1/\epsilon$.

\subsection{Entanglement in mixed states}
\label{ssec:negativity}

Quantum walks with decoherence on the line and the cycle thus provide
two examples where the optimal computational properties are
obtained for a judicious combination of quantum dynamics and decoherence.
In order to investigate this phenomenon in more detail, we asked
how ``quantum'' the system is by the end of the quantum walk when the
optimal amount of decoherence is applied.  To quantify the ``quantumness'', we 
used the entanglement between the coin and the position, which depends on 
the quantum correlations in the system.

We chose our entanglement measure to be the negativity
\cite{peres96a,horodecky96a,vidal01a} because this can 
be calculated numerically in a fairly straightforward manner for density
operators such as $\bm\rho(t)$, and there are few options that meet this
criterion.  First we must choose
a division of our system into two (or more) subsystems between which to
identify the entanglement.  For our quantum walk, the natural division is
between the coin and the quantum walker's position.
We note that the entanglement across this division will be the same
whether we regard the quantum walk as a physical system with a qubit
coin and a unary position, or as an algorithm running on a quantum
computer with a single qubit for the coin and the position encoded
in binary in the remainder of the quantum register.
We perform a partial transpose on one subsystem to obtain a new matrix 
$\bm\rho'(t)$.  For example, the partial transpose with respect to
the coin subsystem is 
\begin{equation}
\rho'_{xc,yb}(t) = \rho_{xb,yc}(t)
\end{equation}
where $x$, $y$ are position indices and $c$, $b$ are coin state indices.
Next, we determine the spectrum of $\bm\rho'(t)$, denoted by $\{\lambda'_i\}$.
The normalisation of $\bm\rho(t)$ is carried over to $\bm\rho'(t)$, so
$\sum_i\lambda'_i = 1$, but unlike $\bm\rho(t)$, it is possible for
$\bm\rho'(t)$ to have negative eigenvalues.  The negativity is defined 
\cite{peres96a,horodecky96a,vidal01a} as
\begin{equation}
E=\frac{1}{2}\left(\sum_i|\lambda'_i| -1\right),
\end{equation}
which is just the sum of the negative eigenvalues.
The negativity ranges between zero and one, with any non-zero value
indicating entanglement is present.  If the negativity is zero, it means the
state is probably not entangled, but there can be exceptions
\cite{horodecky98a}.  The exceptions are known to be relatively rare
in the set of all possible states \cite{zyczkowski99a}, and the
entanglement they contain is difficult to apply to useful quantum tasks
\cite{horodecky98a}.  For this study, we will not need the fine-grained
detail of these possible exceptions.

The entanglement in a pure state quantum walk has been studied previously,
see for example, \cite{carneiro05a}.  It fluctuates with each step, and
eventually settles down to an asymptotic value that depends on the initial
state of the quantum coin, and on any bias in the quantum coin operator
$\mathbf{C}$.  The addition of decoherence smooths out this behaviour, see
\begin{figure}
    \begin{center}
        \resizebox{0.6\columnwidth}{!}{\includegraphics{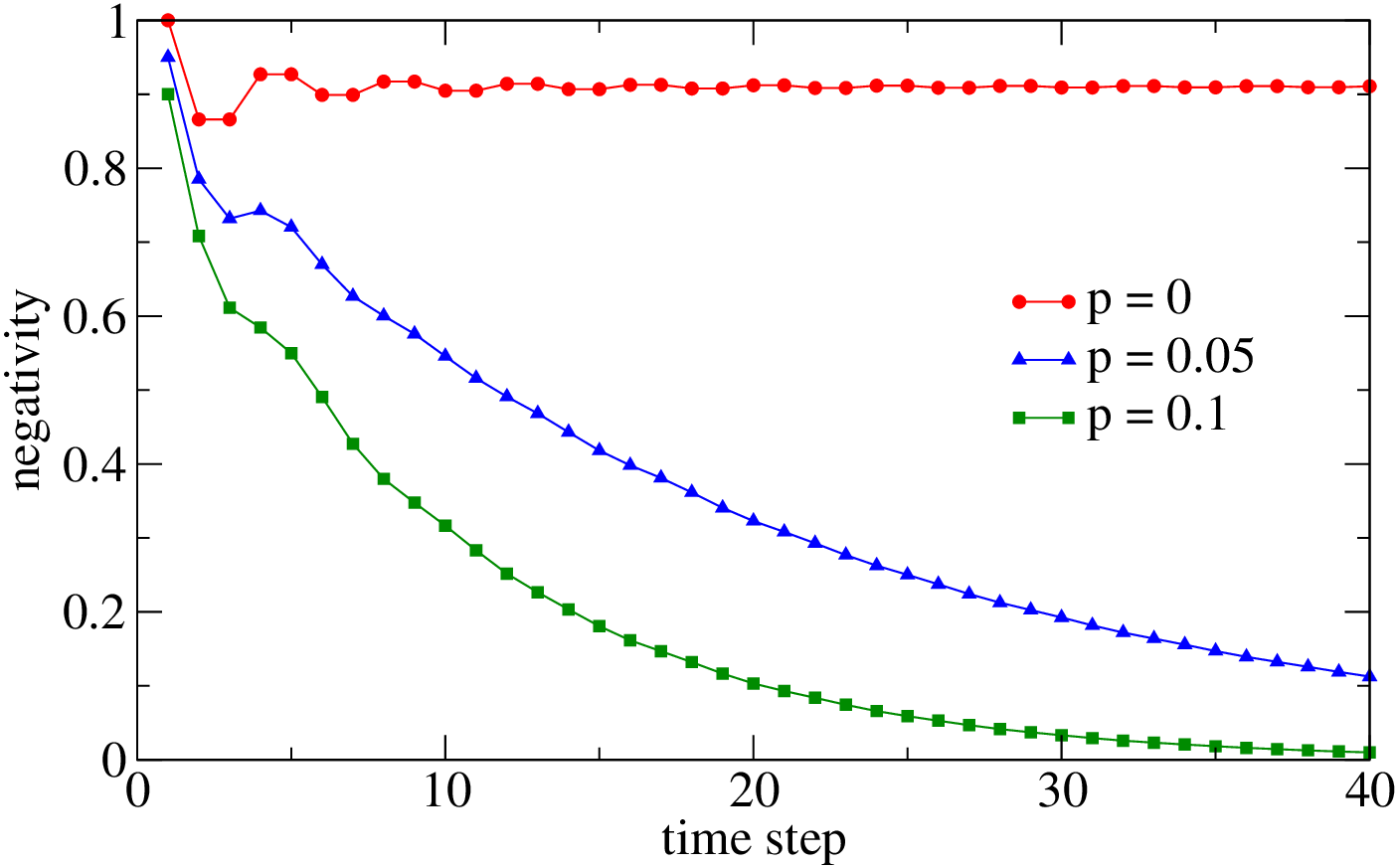}}
    \end{center}
    \caption{Numerical data showing the negativity between the coin and position 
	for decoherence rates $p=0$ (red), $p=0.05$
        (blue) and $p=0.1$ (green) for a quantum walk on the
        line with decoherence on both coin and position,
	run for 40 time steps.}
    \label{fig:ld40neg}
\end{figure}
figure~\ref{fig:ld40neg}, and steadily reduces the level of entanglement
between the coin and the position.

\section{Results and conclusions}
\label{sec:results}

Our simulations simply take equation~(\ref{eq:decdyn}) and evolve it
numerically, calculating the negativity and, using an appropriate
top hat or uniform distribution,
the TVD for various types of decoherence and values of
the decoherence rate $p$.  We studied quantum walks on both the line and
the cycle, for various initial states, lengths of walk, and sizes of
cycles, with decoherence applied to the coin only, the position only, and
to both the coin and position.  
These simulations can be accomplished on a desktop computer using
straightforward code written in a language such as C++.
We used various Apple Mac G4 and
G5 computers and the versions of the GNU C compiler that come with
the operating system OSX 10.4 and associated Developer Tools.
Our results are summarised in the following subsections.

\subsection{Entanglement in the walk on the line}
\label{sec:linent}

\begin{figure}
    \begin{center}
        \resizebox{0.6\columnwidth}{!}{\includegraphics{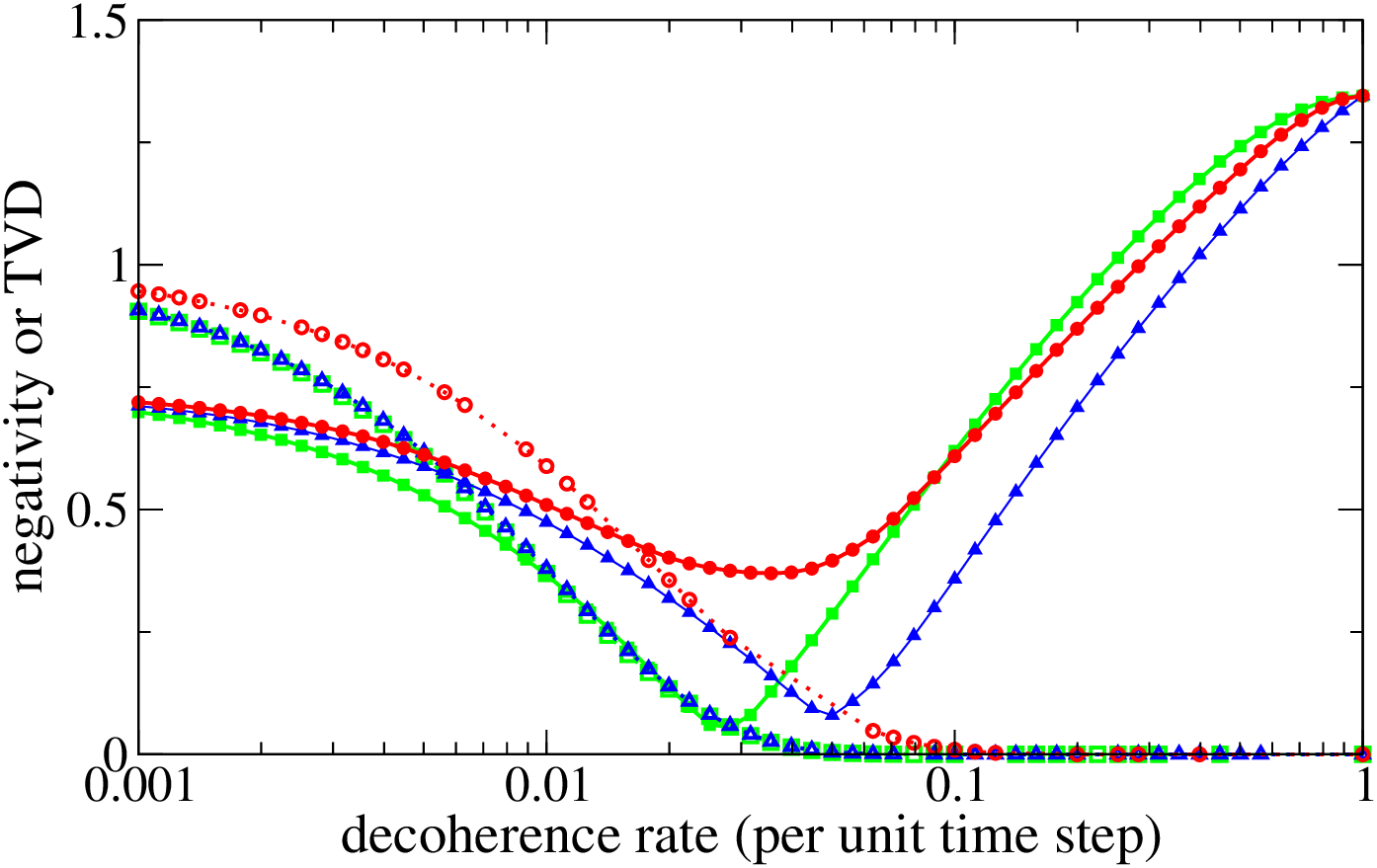}}
    \end{center}
    \caption{Numerical data showing the negativity (dotted, open symbols)
	and TVD (solid, filled symbols)
	for decoherence applied to the position only (blue triangles),
	the coin only (red circles) and to both coin and position
	(green squares) for a quantum walk on the line of 100 steps.
	Some negativity data is missing (missing symbols).}
    \label{fig:ld100}
\end{figure}
Guided by the results in \cite{kendon02c} and \cite{carneiro05a},
we focused on the entanglement at the end of the quantum walk on the line,
just before a final measurement would be made to find out where the quantum
walker has ended up.  We considered how the entanglement varies as
the decoherence rate $p$ is varied.
As well as testing three cases of decoherence, applying it to both the coin
and position, and also separately to just the coin or the position,
we also tested two different starting states for the coin, both of
which produce symmetrical distributions \cite{tregenna03a,konno02b},
these being $(\ket{{-1}}+i\ket{{+1}})/\sqrt{2}$ and
$\cos(\pi/8)\ket{{-1}} + \sin(\pi/8)\ket{{+1}}$.
The two initial states achieve a symmetrical distribution in fundamentally 
different ways, the former by combining two asymmetric distributions that
do not interact, and the latter by arranging the interference exactly the
right way to produce a symmetrical outcome.
The distributions are not identical, and we observed that
the TVD and negativity differ slightly, but not in any significant way.
The negativity calculations require 
diagonalisation of large matrices (to find the eigenvalues), 
and we found for larger matrices (more time steps) our numerical
routines (based on methods from Numerical Recipes \cite{nrc93})
did not always manage to do this.  Failure was always reported
by the routines, and we have simply left gaps in the figures where
the negativity data is missing.
We obtained sufficient results to show the overall trends,
as can be seen in a typical example from our results shown in
figure~\ref{fig:ld100}.
Both the TVD from the optimal top hat distribution,
and the negativity are plotted.  The minimum in the TVD
indicates the optimal decoherence rate $p$.  For a quantum walk of 100 steps
the optimal $p$ is in the range $0.025\lesssim p\lesssim 0.05$.
Note that the minimum for measurements on the coin only is much shallower.
The distribution in this case retains a cusp shape \cite{kendon02c}.
The negativity is also higher than when decoherence is applied
to the position of the walker, indicating that coin decoherence is less
effective at removing the quantum correlations.

If decoherence is applied to the position, with or without decoherence on the
coin as well, the negativity drops to zero at $p\simeq0.055$,
shortly after the optimal decoherence rate is reached.
The optimal amount of decoherence is thus just about the amount required to
remove the the quantum correlations from the system.

\subsection{Entanglement in the walk on a cycle}
\label{ssec:cyclent}

\begin{figure}
    \begin{center}
        \resizebox{0.6\columnwidth}{!}{\includegraphics{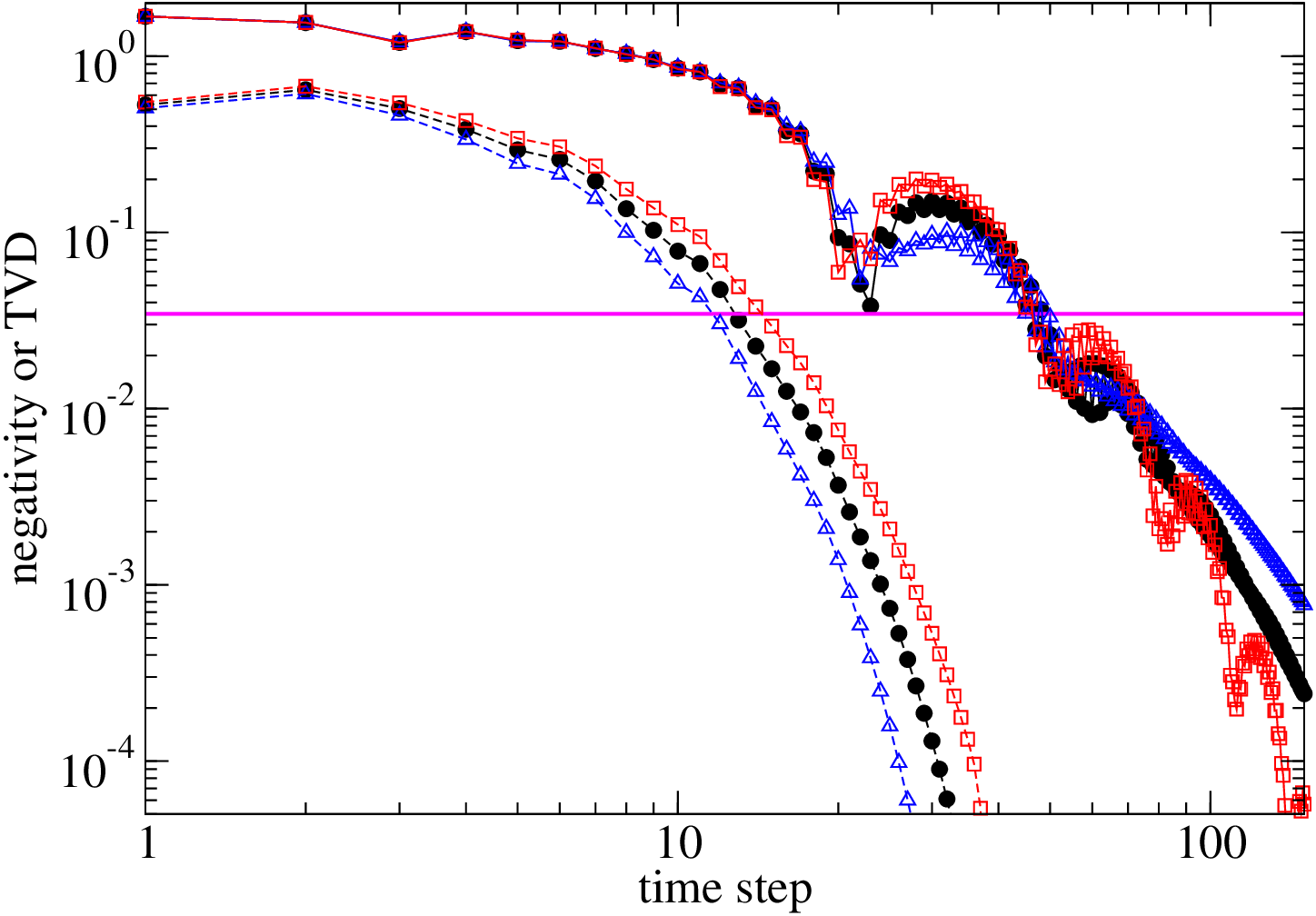}}
    \end{center}
    \caption{Numerical data showing the negativity (dashed) and TVD (solid) for
        decoherence applied to the position for a quantum walk on the
        cycle of size 29 with near-optimal decoherence rates
	$p=0.2239$ (blue triangles),
        $p=0.2511$ (black circles) and $p=0.2818$ (red squares).
        The pink horizontal line is at $1/29$.}
    \label{fig:cd29neg}
\end{figure}
For the mixing time on cycles, there are extra considerations.  The mixing
time depends not only on the size of the cycle, but also on how
close to uniform one sets the threshold $\epsilon$, see equations
(\ref{eq:mixdef}) and (\ref{eq:meanmixdef}.
As already noted, pure quantum walks don't mix unless something is done to
disrupt the pure quantum evolution. Both random decoherence and regular repeated
measurements can efficiently change the behaviour into that of fast mixing
to the uniform distribution \cite{richter06b}.
Again decoherence applied to the coin only does not provide a significant
improvement in the behaviour compared with decoherence applied to the position. 
Note also that the optimal rate of measurement
$2\lesssim pN\lesssim 5$ is independent of the threshold $\epsilon$,
even though the main effect is to provide logarithmic scaling of the
mixing time with $\epsilon$.

We studied how the entanglement varies during the quantum walk
on a cycle, with the decoherence rate $p$ chosen to be near-optimal.
As with the walk on the line, we examined the three cases
of decoherence applied to both the position and coin, and applied to just
the position or coin separately.
The results for a typical example, a cycle of size $N=29$ with decoherence
applied to the position only, are shown in figure~\ref{fig:cd29neg}.
While the actual mixing time is determined by the choice of $\epsilon$, 
we have indicated the position of $1/N$ in figure~\ref{fig:cd29neg},
this being the probability of finding the walker at one location
in a uniform distribution.
The time at which the TVD drops below this line is around the time the
entanglement also drops to zero.  The variability of the TVD shows that the
mixing time is not a smooth function of $\epsilon$, so we cannot expect
to determine a more precise result.  Nonetheless, we have the same
qualitative behaviour as we found for the quantum walk on a line: the
optimal amount of decoherence is just enough to remove all of the quantum
correlations.

\subsection{Discussion and conclusion}
\label{ssec:conc}

For the purpose of uniform sampling, the optimal quantum walk on
the line and the cycle is a carefully balanced combination of
quantum dynamics with decoherence or measurements providing
randomness during the evolution of the walk.
We have shown that, while the quantum correlations are necessary
to obtain linear spreading and mixing times, they must be neutralised
to produce a uniform final distribution.
This make intuitive sense: quantum correlations
distort the distribution, giving it peaks and troughs, especially
at the ends of the top hat \cite{nayak00a,ambainis01a}.
On the other hand, if the decoherence rate is turned up
until the classical random walk is
obtained, classical correlations build up to produce the
binomial distribution in which the quantum walker is more likely to be
found nearer the starting point of the walk.
A uniform distribution, by definition, has neither quantum nor classical
correlations.

The fact that decoherence applied only to the coin does not produce
a good top hat distribution shows that this result is non-trivial.
There is no guarantee that decoherence can be arranged to remove
the quantum correlations almost completely before any significant
classical correlations build up, but for the quantum walk on the line
and the cycle it is possible to do this by applying decoherence to the 
position.  When decoherence is applied to the coin only, classical
correlations have started to build up before all the quantum 
correlations are removed.

Applied to quantum computation, this result has intriguing 
implications.  For the task of uniform sampling, the optimal 
way to use a quantum walk is not a pure quantum process, it is a 
slightly noisy process which has lost all entanglement by the end of
the computation.  Given the difficulty of arranging for perfect 
noiseless quantum operations, it might be easier to implement an optimal 
quantum walk since it doesn't have to be perfect.  It also begs the question
whether there are other computational tasks for which a noisy quantum
process is optimal.  In our search for perfect quantum computation
we may have overlooked the possibility that some useful tasks may not need it.
A communications task in which added randomness
improves the performance is already known: distilling a secret key from 
imperfect quantum exchanges \cite{kraus04a,renner05a}.  Here the added
noise reduces the information the eavesdropper can obtain about the secret key,
and leads to better key rates than a deterministic key distillation process.

\begin{acknowledgments}

We dedicate this paper to Ivens Carneiro, whose contribution to the
work in \cite{carneiro05a} laid the foundations for these results.
We learned with much sadness of his untimely death in a road accident
in April 2006.

We also thank many other people for interesting discussions of quantum walks,
among them,
Hilary Carteret,
Jochen Endrejat,
Barbara Kraus,
Peter Richter,
Barry Sanders,
Mario Szegedy,
and
Tino Tamon
stimulated our thinking for the work in this paper.
VK is funded by a Royal Society University Research Fellowship.

\end{acknowledgments}


\bibliography{../bibs/qrw,../bibs/qit,../bibs/ent,../bibs/misc}



\end{document}